\documentclass[aps,pra,groupedaddress,amsmath,amssymb,showpacs]{revtex4}
\usepackage[dviwin]{graphicx}
\usepackage[dvips]{color}
\usepackage{graphicx}
\usepackage{dcolumn}
\usepackage{bm}
\begin{document}
\title{Qutrit: entanglement dynamics in the finite qutrit chain in the
consistent magnetic field}
\author{E. A. Ivanchenko}
\email{yevgeny@kipt.kharkov.ua}
\affiliation{Institute for
Theoretical Physics, National Science Center
\textquotedblleft{}Institute of Physics and
Technology\textquotedblright{},
 \\
1, Akademicheskaya str., 61108 Kharkov, Ukraine}
\date{\today}
\begin{abstract}
Based on the Liouville-von Neumann equation, we obtain closed
system of equations for the description of a qutrit or coupled
qutrits in arbitrary time-dependent external magnetic field. The
dependence of the dynamics on the initial states and magnetic
field modulation is studied analytically and numerically. We
compare the relative entanglement measure's dynamics in the
bi-qutrit system with permutation particle symmetry. We find the
magnetic field modulation which retains the entanglement in the
system of two coupled qutrits. Analytical formulas for
entanglement measures in the chain from 2 to 6 qutrits are
presented.
\end{abstract}


\maketitle

\section{Introduction}
Multi-level quantum systems are studied intensively, since they
have wide applications.  Some of the existent analytical results
for spin 1 \cite{Hioe} are derived in terms of the coherent vector
\cite{HioeEberly}. The class of exact solutions for a three-level
system is given in Ref. \cite{AMIshkhanyan}. The application of
coupled multi-level systems in quantum devices is actively studied
\cite{ZobovShauroErmilov}. The study of these systems is topical
in view of possible applications for useful work  in microscopic
systems \cite{Scully}. Exact solutions for two uncoupled qutrits
interacting with vacuum are obtained in Ref.
\cite{DerkaczJakobczyk}. For the case of the qutrits interacting
with stochastic magnetic field exact solutions are obtained in
Ref. \cite{MazharAli}. Exact solutions for coupled qutrits in
magnetic field as far as we know were not found.\\
 The entanglement in multi-particle coupled systems is an important
resource for many problems in quantum information science, but its
quantitative value is difficult  because of different types of
entanglement. Multi-dimensional entangled states are interesting
both for the study of the foundations of quantum mechanics and for
the topicality of developing new protocols for quantum
communication. For example, it was shown that for maximally
entangled states of two quantum systems, the qudits break the
local realism stronger than the qubits
\cite{KaszlikowskiGnaciZukowskiMiklaszewskiZeilinger}, and that
the entangled qudits are less influenced by the noise than the
entangled qubits. Using entangled qutrits or qudits instead of
qubits is more protective from interception. From the practical
point of view, it is clear that generating and saving the
entanglement in the controlled manner is the primary problem for
the realization of the quantum computers. The maximally entangled
states are best suited for the protocols of quantum
teleportation and quantum cryptography.\\
 The entanglement and the
symmetry are the basic notions of the quantum mechanics. We study
the dynamics of multipartite systems, which are invariant at any
subsystem permutation. The aim of our work is finding exact
solutions for the dynamics of coupled qutrits interacting with
alternating magnetic field as well as the comparative analysis of
the entanglement measures in the chain of
qutrits.\\
 The rest of the paper is organized as following. The
Hamiltonian of the anisotropic qutrit in arbitrary alternating
magnetic field is described in Sec. II. Then the system of
equations for the description of the qutrit dynamics is derived in
the Bloch vector representation. We introduce the consistent
magnetic field, which describe entire class of field forms. In
section III we derive the system of equations for the description
of the dynamics of two coupled qutrits in the consistent field and
find the analytical solution for the density matrix in the case of
anisotropic interaction. Analytical formulas, which describe the
entanglement in spin chains from 2 to 6 qutrits, are presented in
Sec. IV. The results are demonstrated graphically in Sec. V for
concrete parameters. The brief conclusions are given in Sec. VI.
The auxiliary analytical results are presented in the Appendices.
%
%

\section{Qutrit}\label{qutrit}
\subsection{Qutrit Hamiltonian}\label{Qutrit Hamiltonian}
 We take the qutrit Hamiltonian (for the spin-1
particle) in the space of one qutrit $\mathbb{C}^{3}$ in the basis $%
|1>=(1,0,0),\;|0>=(0,1,0),\;|-1>=(0,0,1)$, in external magnetic field $%
\overrightarrow{\bm{h}}=(h_{1},h_{2},h_{3})$ with anisotropy, in the form
\begin{equation}\label{eq:1}
\hat{H}
=h_{1}S_{1}+h_{2}S_{2}+h_{3}S_{3}+Q(S_{3}^{2}-2/3E)+d(S_{1}^{2}-S_{2}^{2}),
\end{equation}
where $h_{1},\;h_{2},\;h_{3}$ are the Cartesian components of the external
magnetic field in the frequency units (we assume $\hbar =1$, Bohr magneton $\mu_B=1$); $%
S_{1},\;S_{2},\;S_{3}$ are the spin-1 matrices (see Appendix A);
$E$ stands for the $3\times 3$ unity matrix; \thinspace $Q,\;d$
are the anisotropy constants. When the constants $Q,\;d$ are
zeros, then the two Hamiltonian eigenvalues are symmetrically
placed in respect to the zero level.
\subsection{Liouville-von Neumann equation}\label{Decomplexification}
 The qutrit dynamics in the magnetic field we
describe in the density matrix formalism with the Liouville-von Neumann
equation
\begin{equation}\label{eq:2}
i\partial _{t}\rho =[\hat{H},\,\rho ],~\rho (t=0)=\rho _{0}.
\end{equation}
It is convenient to rewrite Eq. \eqref{eq:2} presenting the density matrix $%
\rho $ in the decomposition with the full set of orthogonal Hermitian
matrices $C_{\alpha }$ \cite{Morris} (further the summation over Greek
indices will be from 0 to 8 and over the Latin ones from 1 to 8, see
Appendix A)
\begin{equation} \label{eq:3}
\rho =\frac{1}{\sqrt{6}}C_{\alpha }R_{\alpha }=\left(
\begin{array}{ccc}
\frac{1}{3}+\frac{R_{3}}{\sqrt{6}}+\frac{R_{6}}{\sqrt{18}} &
\frac{
R_{1}+R_{7}-i(R_{2}+R_{5})}{\sqrt{12}} & \frac{-iR_{4}+R_{8}}{\sqrt{6}} \\
\frac{R_{1}+R_{7}+i(R_{2}+R_{5})}{\sqrt{12}} &
\frac{1}{3}-\frac{2R_{6}}{
\sqrt{18}} & \frac{R_{1}-R_{7}-i(R_{2}-R_{5})}{\sqrt{12}} \\
\frac{iR_{4}+R_{8}}{\sqrt{6}} & \frac{R_{1}-R_{7}+i(R_{2}-R_{5})}{\sqrt{12}}
& \frac{1}{3}-\frac{R_{3}}{\sqrt{6}}+\frac{R_{6}}{\sqrt{18}} \\
&  &
\end{array}%
\right).
\end{equation}
Since $\mathrm{Tr\,}C_{i}=0$ for $1\leq i\leq 8$, then from the condition $%
\mathrm{Tr\,}\rho =R_{0}$ it follows that $R_{0}=1$. And although the
results are independent of the basis choice, in this basis the functions $%
R_{i}=\mathrm{Tr\,}\rho \,C_{i}$ have the concrete physical meaning \cite%
{AllardHard}. The values $R_{1},R_{2},R_{3}$ are the polarization vector
Cartesian components; $R_{4}$ is the two-quantum coherence contribution in $%
R_{2}$; $R_{5}$ is the one-quantum anti-phase coherence contribution in $%
R_{2}$; $R_{6}$ is the contribution of the rotation between the
phase and anti-phase one-quantum coherence; $R_{7}$ is the
one-quantum anti-phase coherence contribution in $R_{1}$; $R_{8}$
is the two-quantum coherence contribution in $R_{1}$.\\
Under the unitary evolution the length of the generalized Bloch
vector
\begin{equation}\label{eq:4}
b=\sqrt{R_{i}^{2}}
\end{equation}
is conserved. The length of the generalized vector \eqref{eq:4}
for pure states equals to $\sqrt{2}$. Since $i\partial _{t}\rho
^{n}=[\hat{H},\rho ^{n}]$ $(n=1,2,3,\dots )$, then under unitary
evolution there is countable number of the conservation laws
$\mathrm{Tr\,}\rho =c_{1}=1,~\mathrm{Tr\,} \rho ^{2}=c_{2},\dots
$, from which only $c_{2},\,c_{3}$ are algebraically independent
\cite{Elgin}. Additional quadric invariants of motion can be
easily obtained after equating the matrix elements in defining the
pure state. For example, two of these invariants, which follow
from the expression $(\rho ^{2}-\rho )_{13}=0$, have the form
\begin{equation}\label{eq:5}
R_{1}^{2}-R_{2}^{2}+R_{5}^{2}-R_{7}^{2}-2\sqrt{\frac{2}{3}}(1-\sqrt{2}
R_{6})R_{8}=0,\;R_{5}R_{7}-R_{1}R_{2}+\frac{2}{\sqrt{3}}(\frac{1}{\sqrt{2}}
-R_{6})R_{4}=0.
\end{equation}
For numerical calculations, these invariants control also the signs of the
values $R_{i}$ and thus the using of the invariants is useful when the
analytical solutions are difficult to find. According to the Kelly-Hamilton
theorem, the density matrix $\rho $ satisfies to its characteristic equation%
\begin{equation}\label{eq:6}
\rho ^{3}-\rho ^{2}+\frac{2-b^{2}}{6}\rho -\det \rho \,E=0.
\end{equation}
From equation \eqref{eq:6} it follows that the density matrix
determinant $ \det \rho =(\mathrm{Tr\,}\rho ^{3}-\mathrm{Tr\,}\rho
^{2})/3+(2-b^{2})/18$ is also the motion invariant. The
Liouville-von Neumann equation in terms of the functions $R_{i}$
takes the form of the closed system of 8 real differential
first-order equations. This system of equations in the compact
form can be written as following \cite{Elgin,JMPIvanchenko}:
\begin{equation}\label{eq:7}
\partial _{t}R_{l}=e_{ijl}h_{i}R_{j},
\end{equation}
where $e_{ijl}$ are the antisymmetrical structure constants, $
h_{i}=2(h_{1},h_{2},h_{3},0,0,\frac{Q}{\sqrt{3}},0,d)$ are the
Hamiltonian components \eqref{eq:1} in the basis  $C_\alpha$(see
Appendix A).
\subsection{The consistent field}\label{The consistent field}
Consider the qutrit dynamics in the alternating field of the form
\begin{equation}\label{eq:8}
\vec{h}(t)=\left( \omega _{1}\mathrm{cn}(\omega t|k),\;\omega
_{1}\mathrm{sn} (\omega t|k),\;\omega _{0}\mathrm{dn}(\omega
t|k)\right) ,
\end{equation}
where $\mathrm{cn},\mathrm{sn},\mathrm{dn}$ are the Jacobi
elliptic functions \cite{AbramovitzStegun}. Such field modulation
under the changing of the elliptic modulus $k$ from 0 to 1
describes the whole class of field forms from trigonometric
($\mathrm{cn}(\omega t|0)=\mathrm{cos}\omega
t,\;\mathrm{sn}(\omega t|0)=\mathrm{sin}\omega
t,\;\mathrm{dn}(\omega t|0)=1$ ) \cite{IIRabi} to the
exponentially impulse ones ($\mathrm{cn}(\omega t|1)=
\frac{1}{\mathrm{ch}\omega t},\;\mathrm{sn}(\omega
t|1)=\mathrm{th}\omega t,\;\mathrm{dn}(\omega
t|1)=\frac{1}{\mathrm{ch}\omega t}$) \cite {BambiniBerman}. The
elliptic functions $\mathrm{cn}(\omega t|k)$ and$\;
\mathrm{sn}(\omega t|k)$ have the real period $\frac{4K}{\omega
}$, while the function $\mathrm{dn}(\omega t|k)$ has the two times
smaller period. Here $K$ is the full elliptic integral of the
first kind \cite{AbramovitzStegun}. In other words, even though
the field is periodic with common real period $\frac{4K}{\omega
}$, but as we can see, the frequency of the longitudinal field
amplitude modulation is two times higher than the one of the
transverse
field. Such field we call consistent.\\
\indent Let us make use of the substitution $ \rho =\alpha
_{1}^{-1}r\alpha _{1}$. Then we obtain the equation for the matrix
$r$ in the form
\begin{equation}\label{eq:9}
i\partial _{t}r=[\alpha _{1}\hat{H}\alpha _{1}^{-1}-i\alpha
_{1}\partial _{t}(\alpha _{1}^{-1}),r]
\end{equation}
with the matrix
\begin{equation}\label{eq:10}
\alpha _{1}=\left(
\begin{array}{ccc}
f & 0 & 0 \\
0 & 1 & 0 \\
0 & 0 & f^{-1} \\
\end{array}
\right) ,
\end{equation}
where $f(\omega t|k)=\mathrm{cn}(\omega t|k)+i\mathrm{sn}(\omega t|k).$
Since \newline
\begin{equation}\label{eq:11}
\alpha _{1}S_{1}\alpha _{1}^{-1}=S_{1}\mathrm{cn}(\omega
t|k)-S_{2}\mathrm{sn}(\omega t|k),\;\alpha _{1}S_{2}\alpha
_{1}^{-1}=S_{1}\mathrm{sn}(\omega t|k)+S_{2}\mathrm{cn}(\omega
t|k),\;\alpha _{1}S_{3}\alpha _{1}^{-1}=S_{3},
\end{equation}
then the equation for the matrix $r$ without taking into account the
anisotropy can be written as following
\begin{equation}\label{eq:12}
i\partial _{t}r=[\omega _{1}S_{1}+\delta \,\mathrm{dn}(\omega
t|k)S_{3},r],\;r(t=0)=\rho _{0},\;\delta =\omega _{0}-\omega .
\end{equation}
At $k=0$ equation \eqref{eq:12} describes the dynamics of the
qutrit in the circularly polarized field
\cite{IIRabi,MillerSuitsGarroway,GrifoniHanggi}. The exact
solutions of this equation are known and at some initial
conditions the explicit formulas are given in Ref.
\cite{NathSenGangopadhyay}. At exact resonance, $\omega =\omega
_{0}$ it is straightforward to present Eq. \eqref{eq:2} in the
deformed field ($k\neq0$) \eqref{eq:8} for the given initial
condition $\rho =\rho _{0}$:
\begin{equation}\label{eq:13}
\rho (t)=\alpha _{1}^{-1}e^{-i\omega _{1}t S_{1}}\rho
_{0}e^{i\omega _{1}t S_{1}}\alpha _{1}.
\end{equation}
Explicit solutions for some specific initial conditions are given
in the Appendix B, Eqs. \eqref{eq:B1} -- \eqref{eq:B6}. From the
explicit exact solutions in the deformed field  at resonance
$\delta =0$ one can see that the populations and the transition
probabilities do not depend on the field deformation
(it is independent of the $k$ modulus).\\
Consider the solution of Eq. \eqref{eq:12} far from the resonance
in the form of $\delta $\ power expansion
\begin{equation}\label{eq:14}
r(t)=r^{(0)}(t)+r^{(1)}(t)+\cdots.
\end{equation}
Then we put the expansion \eqref{eq:14} in Eq. \eqref{eq:12} and
equate the same degree terms. As the result we obtain the system
of equations for finding $r^{(l)}(t)$:
\begin{subequations}\label{eq:15a}
\begin{equation}
i\partial _{t}r^{(0)}=\omega _{1}[S_{1},r^{(0)}],
\end{equation}\label{eq:15b}
\begin{equation}
i\partial _{t}r^{(l)}=\omega _{1}[S_{1},r^{(l)}]+\delta
\,\mathrm{dn}(\omega t|k)[S_{3},r^{(l-1)}],\,l=1,2,\,\ldots .
\end{equation}
We multiply Eq. \eqref{eq:15b} to the left by the matrix
$e^{i\omega _{1}tS_{1}}$ and to the right by the matrix
$e^{-i\omega _{1}tS_{1}}$ for formation of the integrating
multiplier \cite{StabilizationIvanchenko}. Now finding the terms
$r^{(l)}$ in the series \eqref{eq:14} is defined by the previous
ones $r^{(l-1)}$ as following
\end{subequations}
\begin{equation}\label{eq:16}
r^{(l)}(t)=-i\delta \int_{0}^{t}dt^{\prime }{e^{i\omega
_{1}(t^{\prime
}-t)S_{1}}}\mathrm{dn}(\omega t^{\prime }|k)[S_{3},r^{(l-1)}(t^{\prime })]{%
e^{-i\omega _{1}(t^{\prime }-t)S_{1}}}.
\end{equation}
\section{Bi-qutrit}\label{Biqutrit}
In the space $\mathbb{C}^{3}\otimes \mathbb{C}^{3}$ the two-qutrit
density matrix can be written in the Bloch representation
\begin{equation}\label{eq:17}
\varrho =\frac{1}{6}R_{\alpha \beta }C_{\alpha }\otimes C_{\beta
},\;R_{00}=1,\;\varrho (t=0)=\varrho _{0},
\end{equation}
where $\otimes $ denotes the direct product. The functions $R_{m0},R_{0m}$
characterise the individual qutrits and functions $R_{mn}$ characterise
their correlations. The length of the generalized Bloch vector $\sqrt{%
R_{\alpha \beta }^{2}-1}$ for pure states equals
$2\sqrt{2}$.\newline Consider the Hamiltonian of the system of two
qutrits with anisotropic and exchange interaction in magnetic
field in the following form
\begin{eqnarray}  \label{eq:18}
H_{2} =(\overrightarrow{h}\overrightarrow{S}%
+Q(S_{3}^{2}-2/3E)+d(S_{1}^{2}-S_{2}^{2}))\otimes E+ \nonumber\\
E\otimes
(\overrightarrow{\bar{h}}\overrightarrow{S}+\bar{Q}(S_{3}^{2}-2/3E)+
\bar{d}(S_{1}^{2}-S_{2}^{2}))+JS_{i}\otimes S_{i}
&=&\frac{1}{2}h_{\alpha \beta }C_{\alpha }\otimes C_{\beta },
\end{eqnarray}
where $\overrightarrow{h}$ and$\;\overrightarrow{\bar{h}}$ are the
magnetic field vectors in frequency units, which operate on the
first and the second qudits respectively, and $J$ is the constant
of isotropic exchange interaction.\\
 The system of equations for
two qutrits takes the real form in terms of the functions
$R_{m0},R_{0m},R_{mn}$ as the closed system of 80 differential
equations \cite{JMPIvanchenko}, supplemented by the initial
conditions
\begin{subequations}\label{eq:19}
\begin{equation}\label{eq:19a}
\partial _{t}R_{m0}=\sqrt{\frac{2}{3}}e_{pim}(h_{p0}R_{i0}+h_{pl}R_{il}),\;
\partial _{t}R_{0m}=\sqrt{\frac{2}{3}}e_{pim}(h_{0p}R_{0i}+h_{lp}R_{li}),
\end{equation}
\begin{equation}\label{eq:19b}
\partial _{t}R_{mn}=e_{pim}\left[ \sqrt{\frac{2}{3}}%
(h_{pn}R_{i0}+h_{p0}R_{in})+g_{rln}h_{pr}R_{il}\right]
+e_{pin}\left[ \sqrt{
\frac{2}{3}}(h_{mp}R_{0i}+h_{0p}R_{mi})+g_{rlm}h_{rp}R_{li}\right]
,
\end{equation}
where by definition
\end{subequations}
\begin{equation}\label{eq:20}
\mathrm{Tr\,}\rho C_{\alpha }\otimes C_{\beta
}=\frac{2}{3}R_{\alpha \beta }
\end{equation}
and
$h_{p0}=\sqrt{6}(\overrightarrow{h},0,0,\frac{Q}{\sqrt{3}},0,d),\,h_{0p}=
\sqrt{6}(\overrightarrow{\bar{h}},0,0,\frac{\bar{Q}}{\sqrt{3}},0,\bar{d}
),\,h_{11}=h_{22}=h_{33}=2J$\thinspace are the Hamiltonian
expansion coefficients in the basis $C_{\alpha }\otimes C_{\beta
}$ (other coefficients equal to zero). In equations \eqref{eq:19}
Latin indices $m,\,n$ take the values from 1 to 8. Numerical
values for the structure constants $
e_{pim},\;g_{rlm}$ are given in Appendix A.\\
 The energy of the
coupled qutrits in terms of the correlation functions has the
following form
\begin{equation}\label{eq:21}
E(t)=\frac{1}{3}(h_{p0}R_{p0}+h_{0p}R_{0p}+\sum_{i=1}^{3}h_{ii}R_{ii}).
\end{equation}
We study the dynamics of two qutrits in the magnetic field
$\overrightarrow{h }=(\omega _{1}\mathrm{cn}(\omega t|k)),\;\omega
_{1}\mathrm{sn}(\omega t|k),\;\omega _{0}\mathrm{dn}(\omega
t|k),\overrightarrow{\bar{h}}=(\varpi _{1}\mathrm{cn}(\varpi
t|k),\;\varpi _{1}\mathrm{sn}(\omega t|k),\;\varpi
_{0}\mathrm{dn}(\omega t|k))$ at the anisotropy constants equal to
0. We transform the matrix density $\varrho =\alpha
_{2}^{-1}r_{2}\alpha _{2}$ with the matrix $\alpha _{2}=\alpha
_{1}\otimes \alpha _{1}$. Then equation for the matrix $r_{2}$
takes the form $i\partial _{t}r_{2}=[\widetilde{H},r_{2}]$ with
the transformed Hamiltonian\newline $\widetilde{H}=\left(
\begin{smallmatrix}
{}J+D\left( -2\omega +\varpi _{0}+\omega _{0}\right) &
\frac{\varpi _{1}}{
\sqrt{2}} & 0 & \frac{\omega _{1}}{\sqrt{2}} & 0 & 0 & 0 & 0 & 0 \\
\frac{\varpi _{1}}{\sqrt{2}} & D\left( \omega _{0}-\omega \right)
& \frac{
\varpi _{1}}{\sqrt{2}} & J & \frac{\omega _{1}}{\sqrt{2}} & 0 & 0 & 0 & 0 \\
0 & \frac{\varpi _{1}}{\sqrt{2}} & D\left( \omega _{0}-\varpi _{0}\right) -J
& 0 & J & \frac{\omega _{1}}{\sqrt{2}} & 0 & 0 & 0 \\
\frac{\omega _{1}}{\sqrt{2}} & J & 0 & D\left( \varpi _{0}-\omega \right) &
\frac{\varpi _{1}}{\sqrt{2}} & 0 & \frac{\omega _{1}}{\sqrt{2}} & 0 & 0 \\
0 & \frac{\omega _{1}}{\sqrt{2}} & J & \frac{\varpi _{1}}{\sqrt{2}} & 0 &
\frac{\varpi _{1}}{\sqrt{2}} & J & \frac{\omega _{1}}{\sqrt{2}} & 0 \\
0 & 0 & \frac{\omega _{1}}{\sqrt{2}} & 0 & \frac{\varpi _{1}}{\sqrt{2}} &
D\left( \omega -\varpi _{0}\right) & 0 & J & \frac{\omega _{1}}{\sqrt{2}} \\
0 & 0 & 0 & \frac{\omega _{1}}{\sqrt{2}} & J & 0 & D\left( \varpi
_{0}-\omega _{0}\right) -J & \frac{\varpi _{1}}{\sqrt{2}} & 0 \\
0 & 0 & 0 & 0 & \frac{\omega _{1}}{\sqrt{2}} & J & \frac{\varpi
_{1}}{\sqrt{2
}} & D\left( \omega -\omega _{0}\right) & \frac{\varpi _{1}}{\sqrt{2}} \\
0 & 0 & 0 & 0 & 0 & \frac{\omega _{1}}{\sqrt{2}} & 0 &
\frac{\varpi _{1}}{ \sqrt{2}} & J+D\left( 2\omega -\varpi
_{0}-\omega _{0}\right)
\end{smallmatrix}
\right) .$ \newline Since $D\overset{\mathrm{def}}{\equiv
}\text{dn}(\omega t|k)|_{k=0}=1$, then the transformed Hamiltonian
$\widetilde{H}$ does not depend on time and the solution for the
density matrix in the circularly polarized field has the
form%
\begin{equation}\label{eq:22}
\varrho (t)=\alpha _{2}^{-1}e^{-i\widetilde{H}t}\varrho
_{0}e^{i\widetilde{H} t}\alpha _{2}|_{k=0}.
\end{equation}%
In the consistent field at resonance $\omega =\varpi _{0}=\omega
_{0}=h$ at equal $\varpi _{1}=\omega _{1}$ the Hamiltonian
eigenvalues equal to $ -2J,-J,J,J-2\omega _{1},-J-\omega
_{1},J-\omega _{1},-J+\omega _{1},J+\omega _{1},J+2\omega _{1}$.
This allows to find the exact solution in the closed form for any
initial condition, since the matrix exponent $e^{i\widetilde{H}
t}$ in this case can be calculated analytically.\newline For
larger number of the qutrits with pairwise isotropic interaction,
the generalization is evident. In the case of interaction of
qudits with different dimensionality, the reduction of the
original system to the system with constant coefficients can be
done by choosing, for example, the
transformation matrix for spin-3/2 and spin-2 in the form%
\begin{equation}\label{eq:23}
\mathrm{diag\,}(
f^{3/2},\,f^{1/2},\,f^{-1/2},\,f^{-3/2}) \otimes \mathrm{diag\,}
(f^{2},\,f,\,1,\,f^{-1},\,f^{-2}).
\end{equation}
However, the Hamiltonian eigenvalues cannot be found in the simple analytic
form because of the lowering the system symmetry.
\section{Entanglement in the qutrits}\label{Entanglement in the qutrits}
\subsection{Entanglement in the bi-qutrit}\label{Entanglement in the bi-qutrit}
 For the initial maximally
entangled state, which is symmetrical at the particle permutation,
\begin{equation}\label{eq:24}
|\psi >=\frac{1}{\sqrt{3}}\sum_{i=-1}^{1}|i>\otimes |i>,
\end{equation}
in the consistent field at the resonance $\omega =\varpi
_{0}=\omega _{0}=h$ at equal $\varpi _{1}=\omega _{1},$ the exact
solution for the correlation functions is given in Appendix C. The
correlation functions have the property $R_{\alpha \beta
}=R_{\beta \alpha }$, i.e. the symmetry is conserved during the
evolution, since the initial state and Hamiltonian are symmetric
in respect to the particle permutation.\newline Given the exact
solution, one can find the negative eigenvalues of the partly
transposed matrix $\varrho ^{pt}=(T\otimes E)\varrho $\ (here $T$
denotes the transposition):
\begin{equation}\label{eq:25}
\epsilon _{1}=\epsilon _{2}=-\frac{1}{27}\sqrt{69+28\cos
3Jt-16\cos 6Jt} ,\;\epsilon _{3}=-\frac{1}{27}\left( 5+4\cos
3Jt\right).
\end{equation}
The absolute value of the sum of these eigenvalues $
m_{VW}=|\epsilon _{1}+\epsilon _{1}+\epsilon _{3}| $ defines the
entanglement measure (negativity) between the qutrits
\cite{VidalWerner}.\\
 The entanglement between the qutrits can be
described quantitatively with the measure \cite{SchlienzMahler}
\begin{equation}\label{eq:26}
m_{SM}=\sqrt{\frac{1}{8}(R_{ij}-R_{i0}R_{0j})^{2}}.
\end{equation}%
This measure equals to 0 for the separable state and to 1 for the
maximally entangled state, and it is applicable for both pure and
mixed states.\\
 That is why for the maximally entangled initial
state of two qutrits, the entanglement in the consistent field is
defined by the formulae with the found solution for the density
matrix
\begin{equation}\label{eq:27}
m_{SM}=\frac{1}{\sqrt{6561}}\sqrt{4457+2776\cos 3Jt-632\cos
6Jt-56\cos 9Jt+16\cos 12Jt}.
\end{equation}
This measure is numerically equivalent to the measure $m_{VW}$
\cite {VidalWerner,PermutaionalSymTothGuhne}, which is defined by
the absolute value of the sum of the negative eigenvalues
\eqref{eq:25} of the partly transposed matrix.\\
 According to the
definition \cite{PanLiuLuDraayer} for $2$-qutrit pure state, the
entanglement measure equals to
\begin{equation} \label{eq:28}
\eta _{2}=\frac{1}{2}\sum_{i=1}^{2}S_{i},
\end{equation}%
where $S_{i}=-\mathrm{Tr\,}\rho _{i}\log _{3}\rho _{i}$ is the
reduced von Neumann entropy, the index $i$ numerates the
particles, i.e. the other  particle are traced out.\\
 Since
the qutrit reduced matrix eigenvalues equal to $\lambda
_{1}=\lambda _{2}=\frac{1}{27}(5+4\cos 3Jt),\;\lambda
_{3}=\frac{1}{27}(17-8\cos 3Jt),$ then the entanglement measure in
the bi-qutrit takes the form
\begin{equation}\label{eq:29}
\eta _{2}=-\sum_{i=1}^{3}\lambda _{i}\log _{3}\lambda _{i}.
\end{equation}
Normalized by the unity, the measure I-concurrence, which is easy to
calculate, is defined by the formulae \cite{Mintert}
\begin{equation}\label{eq:30}
m_{I}=\frac{\sqrt{3}}{2}\sqrt{2(1-\mathrm{Tr\,\rho
_{1}^{2})}}=\frac{1}{9} \sqrt{57+32\cos 3Jt-8\cos 6Jt},
\end{equation}
where $\rho _{1}=\frac{1}{\sqrt{6}}C_{\alpha }R_{\alpha 0}$ is the
reduced qutrit matrix.\\
The time-dependence of the measure $m_{SM}$ for the symmetrical
initial state
\begin{equation}\label{eq:31}
|s>=\frac{1}{\sqrt{12}}\sum_{i\neq j}(|i>\otimes |j>+|j>\otimes
|i>)
\end{equation}
takes the form
\begin{equation}\label{eq:32}
m_{SM}^{|s>}=\frac{1}{\sqrt{209952}}\sqrt{102679+19136\cos
3Jt+29312\cos 6Jt-1024\cos 9Jt+800\cos 12Jt};
\end{equation}
at $t=0$ this measure equals to $\sqrt{23/32}$.\\
The measures $m_{VW},\;m_{SM},\;\eta _{2},\;m_{I},\;m_{SM}^{|s>}$
do not depend on the parameters of the consistent field, sign of
the exchange constant at zero anisotropy parameters. It should be
noted that the Wooters entanglement measure (concurrence) in the
system of two qubits with the isotropic interaction in the
circularly polarized field at resonance is also independent of the
alternating  field amplitude
\cite{ShiXunZhangQinShengZhuXioYuKuang}, but depends on the
exchange constant magnitude and the initial conditions only.\\
 At zero external field the entanglement measure \eqref{eq:24}
takes the analytic form at equal non-zero anisotropy parameters
$Q=d=\overline{d}= \overline{Q}$
\begin{equation}\label{eq:33}
m_{SM}(Q)=\frac{1}{\left( 9J^{2}+8QJ+16Q^{2}\right) ^{2}}\sqrt{
\sum_{k=0}^{4}q_{k}\cos \left(
k\sqrt{9J^{2}+8QJ+16Q^{2}}\,t\right)},
\end{equation}
where $
q_{0}=4457J^{8}+11616QJ^{7}+47392Q^{2}J^{6}+85888Q^{3}J^{5}+163072Q^{4}J^{4}
+194560Q^{5}J^{3}+221184Q^{6}J^{2}+131072Q^{7}J+65536Q^{8}
$;\thinspace\ $q_{1}=8J^{2}(J+2Q)^{2}\left(
347J^{4}+518QJ^{3}+1440Q^{2}J^{2}+1504Q^{3}J+1024Q^{4}\right) $;\\
$q_{2}=-8J^{2}(J+2Q)^{2}\left(
79J^{4}+76QJ^{3}+320Q^{2}J^{2}+448Q^{3}J+256Q^{4}\right) $;\\
$q_{3}=-8J^{3}(7J-4Q)(J+2Q)^{3}(J+4Q)$,\
$q_{4}=16J^{4}(J+2Q)^{4}$.
\subsection{Entanglement in the chain of qutrits}\label{Entanglement in the chain of qutrits}
\begin{figure}
\includegraphics[width=2.3in]{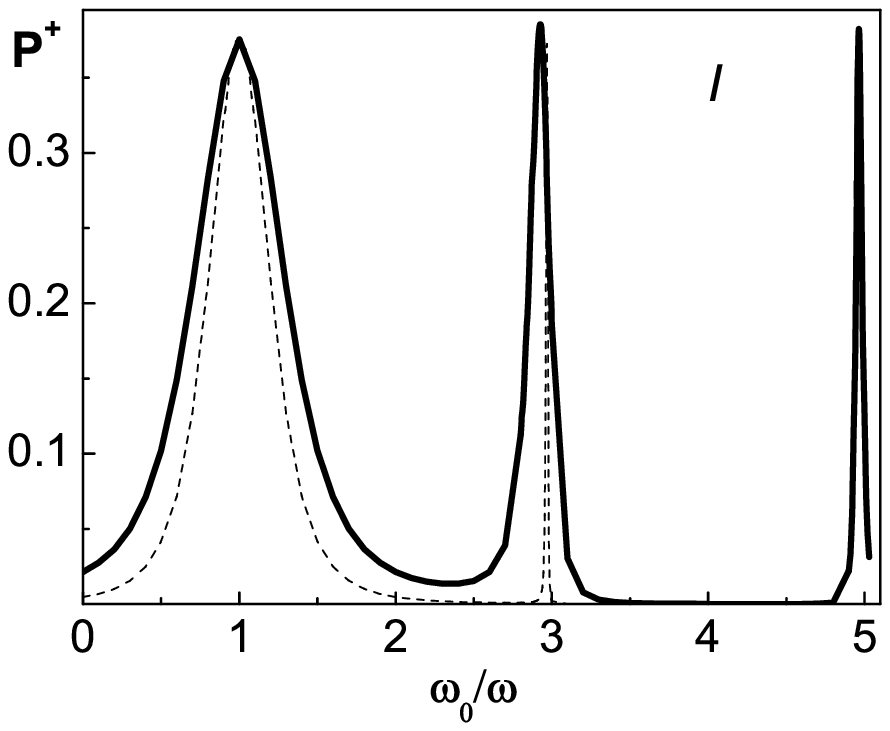} 
\includegraphics[width=2.3in]{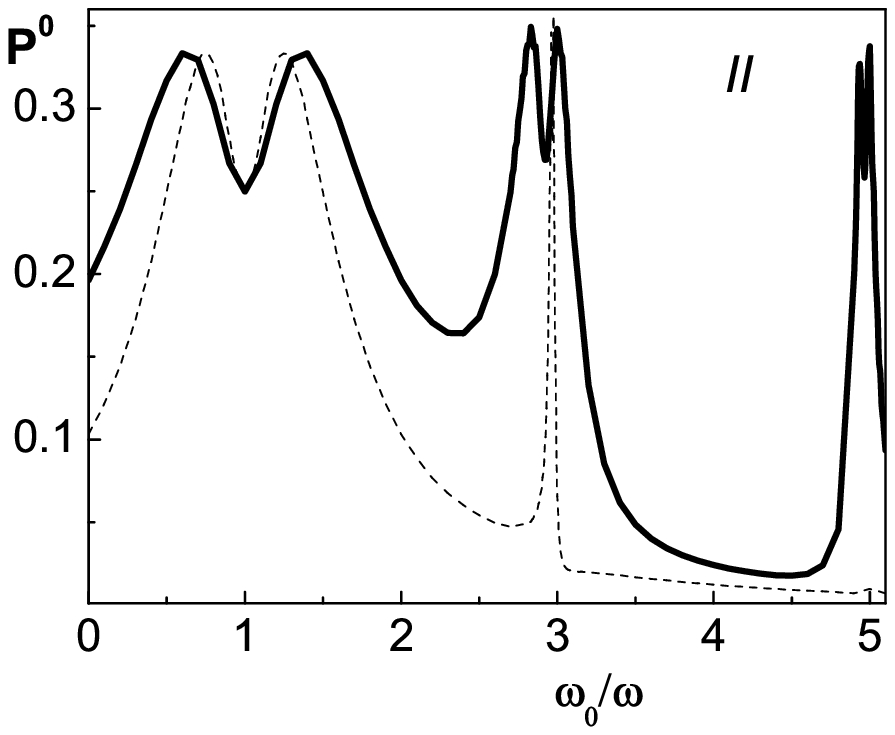} 
\caption{\label{k85N}The time-averaged populations for the initial
pure state $|-1>$ versus the normalized Larmor frequency
$\protect\omega _{0}/\omega $ at the parameters $k=0.85$ (solid
line), $k=0.2$ (dashed line), $d=Q=0$,\thinspace\ $\omega
_{1}=1/3,\; \protect\omega =1$ ($I$ shows the upper level $|1>$
population;\thinspace\ $II$ shows the middle level $|0>$
population).}
\end{figure}
We consider the Hamiltonian of the chain of $N$ qutrits with the pairwise
isotropic interaction in the magnetic field $\overrightarrow{\omega }$ in
the following form
\begin{equation}\label{eq:34}
H_{N}=\sum (\overrightarrow{\omega }\overrightarrow{S}\otimes
\overbrace{ E\otimes \dots \otimes
E}^{N-1}+J\overrightarrow{S}\otimes \overrightarrow{S} \otimes
\overbrace{E\otimes \dots \otimes E}^{N-2}),
\end{equation}%
where the summation is over different possible positions of
$\overrightarrow{ S}$ in the direct products. Because the
maximally entangled state of $N$
qutrits%
\begin{equation}\label{eq:35}
|\phi >=\frac{1}{\sqrt{3}}\sum_{i=-1}^{1}|i>^{\otimes N}
\end{equation}
and the Hamiltonian \eqref{eq:34} have the permutation symmetry,
it follows that the density matrix of $N$ qutrits has the
symmetric correlation functions. The length of the generalized
Bloch vector for pure states equals $\sqrt{3^{N}-1}$.\\
The entanglement measures for the many-particle multi-level
quantum systems are not studied enough and difficult to calculate
in the analytic form, that is why we will present only analytic
formulas for the entropy measure $\eta _{N}$
\cite{PanLiuLuDraayer}, which is defined by the eigenvalues of the
reduced one-particle matrices for each qutrit. As the result of
the mentioned symmetry, the reduced matrices are equal to each
other. Therefore the entanglement measure for $N$ qutrits is
defined by the formulae
\begin{equation}\label{eq:36}
\eta _{N}=-\sum_{i=1}^{3}r_{i}\log _{3}{r_{i}}.
\end{equation}
The eigenvalues of the reduced matrices for 3, 4, 5, and 6 qutrits are
presented in the table below
\begin{equation}\label{eq:37}
\begin{array}{ccc}
N\setminus r_{i} & r_{1}=r_{2} & r_{3} \\
&  &  \\
3 & \frac{29-4\cos 5Jt}{75} & \frac{17+8\cos 5Jt}{75} \\
4 & \frac{905-98\cos 3Jt-72\cos 7Jt}{2205} & \frac{395+196\cos 3Jt+144\cos
7Jt}{2205} \\
5 & \frac{16919-1944\cos 5Jt-800\cos 9Jt}{42525} & \frac{8687+3888\cos
5Jt+1600\cos 9Jt}{42525} \\
6 & \frac{21977-1694\cos 3Jt-1936\cos 7Jt-560\cos 11Jt}{53361} &
\frac{
9407+3388\cos 3Jt+3872\cos 7Jt+1120\cos 11Jt}{53361}. \\
&  &
\end{array}%
\end{equation}
The \ measures $\eta _{3},\;\eta _{4},\;\eta _{5},\;\eta _{6}$
 do not depend on sign of the exchange constant like the
measure $\eta _{2}$.
\begin{figure}
\includegraphics[width=2.3in]{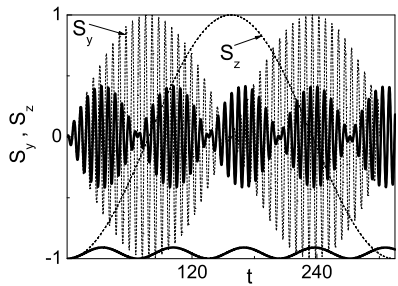}
\caption{\label{SxSz}Dynamics of the spin vector components
$S_{y},\;S_{z}$ for the initial pure state $|-1>$ (dashed lines)
in the circularly polarized field with the parameters:
$k=0,\;\omega _{1}=0.02,\;\omega =\omega _{0}=1,\;d=Q=0$. Solid
lines demonstrate the deformation of the spin components due to
the influence of the second spin (the fluctuator) with $J=0.1$ for
the initial pure state $ |-1>\otimes |-1>$.}
\end{figure}
\section{Numerical results}\label{Numerical results}
\begin{figure}
\includegraphics[width=2.3in]{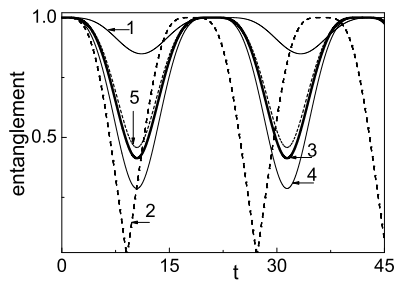}
\caption{\label{EntMSJmJpEntropy}Disentanglement dynamics of the
initially maximally entangled state in the bi-qutrit: in the zero
external field with equal anisotropy constants $
Q=d=\overline{d}=\overline{Q}=0.02507,\,J=-0.1$ (curve 1) and for
$J=0.1$ (curve 2);\thinspace in the consistent field the curve 3
(thick line) demonstrates complete coincidence of the measures
$m_{VW}$ and $m_{SM}$ at $ J $=0.1 and zero anisotropy constants;
the curve 4 demonstrates the entropy measure $\eta _{2}$;
$I$-concurrence is presented by the curve 5 at\ $J$=0.1.}
\end{figure}

\begin{figure}
\includegraphics[width=2.3in]{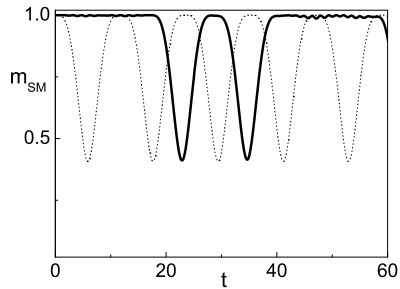}
\caption{\label{Entanglement}Disentanglement of the maximally
entangled state \eqref{eq:24} (solid line) in the impulse field
$\omega _{1}=\varpi _{1}=0,\,\omega _{0}=-\varpi
_{0}=2,\,J=0.178$. The dashed curve presents the evolution in zero
external field.}
\end{figure}

\begin{figure}
\includegraphics[width=2.3in]{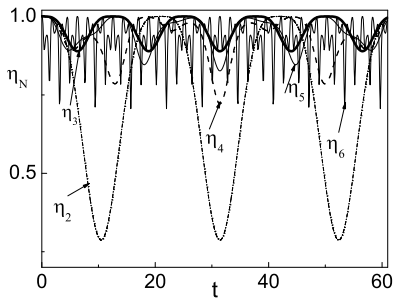}
\caption{\label{h2h3h4h5h6}Disentanglement of the maximally
entangled state \eqref{eq:28} in the chain of 2,\thinspace
3,\thinspace 4,\thinspace 5 and 6 qutrits with $J=0.1$.}
\end{figure}

In Fig. \ref{k85N} we present the populations of the upper and
middle levels in the qutrit averaged over the time interval $ \tau
\rightarrow \infty $: $P^{+}=\frac{1}{\tau }\int_{0}^{\tau
}\,dt\left(
\frac{1}{3}+\frac{1}{\sqrt{6}}R_{3}(t)+\frac{1}{3\sqrt{2}}R_{6}(t)\right)
$ ,\thinspace $P^{0}=\frac{1}{\tau }\int_{0}^{\tau
}\,dt(\frac{1}{3}-\frac{1}{3 }\sqrt{2}R_{6}(t))$ in dependence on
the normalized Larmor frequency $\omega _{0}/\omega $. The
population of the upper level in qutrit coincides in form with the
upper level occupation in a two-level system
\cite{StabilizationIvanchenko}, i.e. this demonstrates the
magnetic resonance position stabilization and the presence of the
parametric resonances.\\
In Fig. \ref{SxSz} we note the considerable suppression of the
qutrit spin oscillations $S_{y}=\text{cn}(\omega t|k)\sin \omega
_{1}t$ and$ \;S_{z}=-\cos \omega _{1}t$ by the environment
(fluctuator) in the case of the resonance $\omega =\omega _{0}$,\
$\overrightarrow{\varpi }=0$.\\
 The bi-qutrit energy \eqref{eq:21}
in the consistent field at isotropic interaction in the case of
the solution \eqref{eq:45}
is constant and equal to $\frac{2}{3}J$.\\
Although the analytic expressions for the measures in a bi-qutrit
$ m_{VW},\;m_{SM}$ are different, but the numerical values are
practically identical. Maximal deviation in the rectangle $(1\geq
J\geq 0.01)\times (100\geq t\geq 0)$ equals $0.014$, where $\times
$ denotes the Cartesian product.\\
Measures $\eta _{2}$ and $m_{I}$ qualitatively coincide with the
measures $
m_{VW},\;m_{SM}$. \\
We have found that the anysotropy of the qutrits disentangles
them, namely the entanglement is decreased down to 0.0010 (see
graphs 1 and 2 in Fig. \ref{EntMSJmJpEntropy}).\\
In the constant longitudinal field $\overrightarrow{\omega }=-
\overrightarrow{\varpi }=(0,\,0,\,\omega _{0})$ (the bi-qutrit
Hamiltonian eigenvalues are equal to
$J,J,x_{1},x_{2},x_{3},-p,-p,p,p$, where $ x_{1},x_{2},x_{3}$ are
the roots of the equation ${
x^{3}+2x^{2}J-p^{2}x-2J^{3}=0,\,p=\sqrt{J^{2}+\omega _{0}^{2}}})$
the Hamiltonian contains the antisymmetric part, thus it follows
that the density matrix for the initial symmetric state will not
be symmetric because of the breaking the symmetry of the particle
permutations. The analytic solution is cumbersome. In the constant
longitudinal impulse field $\overrightarrow{\omega
}=-\overrightarrow{\varpi }=(0,\,0,\,2\,(\theta
((t-17)(t-60))+\theta ((40-t)(57-t)(t-60))))$ the entanglement
dynamics is blocked at $\omega _{0}\gg J$ (Fig. \ref{Entanglement}
). This points to the
possibility to control the entanglement.\\
In Fig. \ref{h2h3h4h5h6} we present the comparative dynamics of
the entropy entanglement measure for 2 to 6 qutrits. The
disentanglement dynamics of the measures $\eta _{3},\eta _{4},\eta
_{5},\eta _{6}$ is similar to the one in the case of two qutrits,
but with smaller oscillation amplitude, i.e. larger number of the
qutrits disentangles less than two qutrits ($0.889\leq \eta
_{3}\leq 1$).
\section{Conclusion}
\label{Conclusion}
 We have shown that the time-averaged
upper-level occupation probability for the qutrit in the
consistent field in dependence on the normalized Larmor frequency
$\omega _{0}/\omega $ coincides in form with the upper-level
occupation probability in the two-level system and the parametric
resonances appear (Fig. \ref{k85N}). In the qutrit coupled to
another qutrit
(fluctuator), the spin oscillations are essentially suppressed.\\
The comparative analysis of the bi-qutrit entanglement measures on
the base of the analytic solution for the density matrix
demonstrates that, in spite of the different approaches to the
derivation of the formulas for the entanglement, all the formulas
give quite close results (Fig. \ref{EntMSJmJpEntropy}), and the
measures $m_{VW}\ $and$\;m_{SM}$ are practically equal. This is in
accordance with the general results for the entanglement in the
systems with the permutational symmetry \cite
{PermutaionalSymTothGuhne}.\\
The analytical formulas for the entanglement measures $\eta
_{3},\eta _{4},\eta _{5},\eta _{6}$ are similar to the
disentanglement measure for two qutrits $\eta _{2}$, but with
numerically smaller oscillation amplitude, i.e. the larger number
of the qutrits disentangles less than two qutrits.\\
\begin{acknowledgments}
The author is grateful to A.~A. Zippa  for fruitful discussions
and constant invaluable support. Many thanks are due to S.~N.
Shevchenko  for help in editing and useful  comments.
\end{acknowledgments}
\section{Appendix A}\label{Appendix A}
The matrix representation of the full set of Hermitian
orthogonal operators for spin-1 has the form
\begin{subequations}
\begin{equation}
C_{1}=S_{1}=\frac{1}{\sqrt{2}}\left(
\begin{array}{ccc}
0 & 1 & 0 \\
1 & 0 & 1 \\
0 & 1 & 0 \\
&  &
\end{array}%
\right) ,\,~C_{2}=S_{2}=\frac{i}{\sqrt{2}}\left(
\begin{array}{ccc}
0 & -1 & 0 \\
1 & 0 & -1 \\
0 & 1 & 0 \\
&  &
\end{array}%
\right) ,\,C_{3}=S_{3}=\left(
\begin{array}{ccc}
1 & 0 & 0 \\
0 & 0 & 0 \\
0 & 0 & -1 \\
&  &
\end{array}%
\right) ,  \label{A1a}
\end{equation}%
\begin{equation}
C_{4}=i\left(
\begin{array}{ccc}
0 & 0 & -1 \\
0 & 0 & 0 \\
1 & 0 & 0 \\
&  &
\end{array}%
\right) ,C_{5}=\frac{i}{\sqrt{2}}\left(
\begin{array}{ccc}
0 & -1 & 0 \\
1 & 0 & 1 \\
0 & -1 & 0 \\
&  &
\end{array}%
\right) ,\,C_{6}=\sqrt{3}(S_{3}^{2}-2/3E)=\frac{1}{\sqrt{3}}\left(
\begin{array}{ccc}
1 & 0 & 0 \\
0 & -2 & 0 \\
0 & 0 & 1 \\
&  &
\end{array}%
\right) ,  \label{A1b}
\end{equation}%
{}
\begin{equation}
C_{7}=\frac{1}{\sqrt{2}}\left(
\begin{array}{ccc}
0 & 1 & 0 \\
1 & 0 & -1 \\
0 & -1 & 0 \\
&  &
\end{array}%
\right) ,\,C_{8}=S_{1}^{2}-S_{2}^{2}=\left(
\begin{array}{ccc}
0 & 0 & 1 \\
0 & 0 & 0 \\
1 & 0 & 0 \\
&  &
\end{array}%
\right) ,\,C_{0}=\sqrt{\frac{2}{3}}E,  \label{A1c}
\end{equation}%
where $E$ is    the unity $3 \times 3 $  matrix. These matrices
have the property of the trace equal to zero ${\rm Tr\,}C_a=0$ and
orthogonality $ {\rm Tr\,} C_a C_b =2 \delta_ {ab} $, $ 1 \leq a,
b\leq 8$. The connection between the basis $C_a$ and the Gell-Mann
basis
${\lambda_a}$ is the following:\\
$C_1=\frac{1}{\sqrt{2}}(\lambda_1+\lambda_6),\;
C_2=\frac{1}{\sqrt{2}}(\lambda_2+\lambda_7),\;
C_3=\frac{1}{2}\lambda_3+\frac{\sqrt{3}}{2}\lambda_8,
C_4=\lambda_5,\; C_5=\frac{1}{\sqrt{2}}(\lambda_2-\lambda_7),\;
C_6=\frac{\sqrt{3}}{2}\lambda_3-\frac{1}{2}\lambda_8,\;
C_7=\frac{1}{\sqrt{2}}(\lambda_1-\lambda_6),\; C_8=\lambda_4.$\\
Non-zero structure constants $e_{abc}$ ($g_{abc}$) antisymmetric
(symmetric) in respect to the permutation of any pair of indices
for the commutators $[C_a,C_b]=2ie_{abc}C_c$ (anticommutators $
\{C_a,C_b\}=\frac{4}{3} E \delta_{ab}+2 g_{abc}C_c $)
 are respectively equal according to the definitions  $e_{abc}=\frac{1}{4i}{\rm Tr\,}[C_a,C_b]C_c$:\\
$e_{123}=e_{147}=e_{158}=-e_{245}=e_{278}=-e_{357}=\frac{1}{2},\;
e_{156}=e_{267}=\frac{\sqrt{3}}{2},\;
e_{348}=-1$;\;$g_{abc}=\frac{1}{4}{\rm Tr\,}\{C_a,C_b\}C_c$:
$g_{336}=g_{446}=-g_{666}=g_{688}=\frac{1}{\sqrt{3}},\;
g_{116}=g_{226}=g_{556}=g_{677}=-\frac{1}{2\sqrt{3}},\;
g_{118}=g_{124}=g_{137}=-g_{228}=g_{235}=-g_{457}=g_{558}=-g_{778}=\frac{1}{2}$.
Hence, the product of the generators is equal to
$C_aC_b=\frac{2}{3}E\delta_{ab}+(g_{abc}+ie_{abc})C_c$.
\section{Appendix B}\label{Appendix B}
 For the initial state $|-1>$ at non-zero detuning $\delta
=\omega _{0}-\omega $ the density matrix elements in the
circularly polarized field have the form
\end{subequations}
\begin{equation} \label{eq:B1}
\rho _{11}=\frac{\omega _{1}^{4}}{\Omega ^{4}}\sin
^{4}\frac{\text{$\Omega $t }}{2},\,\rho
_{12}=-\frac{\sqrt{2}\omega _{1}^{3}}{\Omega ^{4}}\sin ^{3}
\frac{\Omega t}{2}e^{-i\omega t}\left( i\Omega \cos \frac{\Omega
t}{2} +\delta \sin \frac{\Omega t}{2}\right) ,
\end{equation}
\begin{equation}\label{eq:B2}
\rho _{13}=-\frac{\omega _{1}^{2}}{2\Omega ^{4}}\sin
^{2}\frac{\Omega t}{2} e^{-2i\omega t}\left( \omega
_{1}^{2}-2i\delta \Omega \sin \Omega t+\left( 2\delta ^{2}+\omega
_{1}^{2}\right) \cos \Omega t\right) ,
\end{equation}
\begin{equation}\label{eq:B3}
\rho _{22}=\frac{\omega _{1}^{2}\sin ^{2}\frac{\Omega
t}{2}}{\Omega ^{4}} \left( 2\delta ^{2}+\omega _{1}^{2}(1+\cos
\Omega t)\right) ,\,\rho _{23}=- \frac{\omega _{1}}{\sqrt{2}\Omega
^{4}}e^{-i\omega t}\left( 2\delta ^{2}+\omega _{1}^{2}(1+\cos
\Omega t)\right) \left( \delta \sin ^{2}\frac{ \Omega
t}{2}+i\frac{\Omega }{2}\sin \Omega t\right) ,
\end{equation}
\begin{equation}\label{eq:B4}
\rho _{33}=\frac{1}{4\Omega ^{4}}\left( 2\delta ^{2}+\omega
_{1}^{2}(1+\cos \Omega t)\right) ^{2},\,\rho _{ik}=\rho
_{ki}^{\ast },
\end{equation}
where $\Omega =\sqrt{\omega _{1}^{2}+\delta ^{2}}$ is the Rabi
frequency.
\newline
For the initial doubly stochastic state
$\frac{1}{\sqrt{3}}(|-1>+|0>+|1>)$ and for the states $|0>$,\
$\frac{1}{2}|-1>+\frac{1}{\sqrt{2}}|0>+\frac{1}{2} |1>|$ at exact
resonance $\delta =0$ in the consistent field, the density
matrices are respectively equal
\begin{equation}\label{eq:B5}
\left(
\begin{array}{lll}
\frac{1}{12}\left( \cos 2\omega _{1}t+3\right)  &
\frac{1}{12}f^{-1}\left( i \sqrt{2}\sin 2\omega _{1}t+4\right)  &
\frac{1}{12}f^{-2}\left( \cos 2\omega
_{1}t+3\right)  \\
\frac{1}{12}f\left( 4-i\sqrt{2}\sin 2\omega _{1}t\right)  &
\frac{1}{6} \left( 3-\cos 2\omega _{1}t\right)  &
\frac{1}{12}f^{-1}\left( 4-i\sqrt{2}
\sin 2\omega _{1}t\right)  \\
\frac{1}{12}f^{2}\left( \cos 2\omega _{1}t+3\right)  &
\frac{1}{12}f\left( i \sqrt{2}\sin 2\omega _{1}t+4\right)  &
\frac{1}{12}\left( \cos 2\omega _{1}t+3\right)
\end{array}%
\right) ,
\end{equation}
\begin{equation}
\left(
\begin{array}{lll}
\frac{1}{2}\sin ^{2}\omega _{1}t & -\frac{if^{-1}\sin 2\omega
_{1}t}{2\sqrt{2
}} & \frac{1}{2}f^{-2}\sin ^{2}\omega _{1}t \\
\frac{if\sin 2\omega _{1}t}{2\sqrt{2}} & \cos ^{2}\omega _{1}t &
\frac{
if^{-1}\sin 2\omega _{1}t}{2\sqrt{2}} \\
\frac{1}{2}f^{2}\sin ^{2}\omega _{1}t & -\frac{if\sin 2\omega
_{1}t}{2\sqrt{2 }} & \frac{1}{2}\sin ^{2}\omega _{1}t
\end{array}
\right) ,\,\left(
\begin{array}{lll}\label{eq:B6}
\frac{1}{16}\left( 5-\cos 2\omega _{1}t\right)  &
-\frac{if^{-1}\sin 2\omega
_{1}t}{8\sqrt{2}} & \frac{1}{8}f^{-2}\sin ^{2}\omega _{1}t \\
\frac{if\sin 2\omega _{1}t}{8\sqrt{2}} & \frac{1}{8}\left( \cos 2\omega
_{1}t+3\right)  & \frac{if^{-1}\sin 2\omega _{1}t}{8\sqrt{2}} \\
\frac{1}{8}f^{2}\sin ^{2}\omega _{1}t & -\frac{if\sin 2\omega
_{1}t}{8\sqrt{2 }} & \frac{1}{16}\left( 5-\cos 2\omega
_{1}t\right)
\end{array}
\right).
\end{equation}
For both the initial middle level and the doubly stochastic pure
initial state, the populations of the upper and bottom levels are
equal. \cite{NathSenGangopadhyay}. This property is fulfilled for
the mixed state as well.
\section{Appendix C}\label{Appendix C}
The exact solution for the correlation functions of the
initial state \eqref{eq:24}, which is symmetric under the particle
permutation, in the consistent field at resonance $\omega =\varpi
_{0}=\omega _{0}=h$ and equal $\varpi _{1}=\omega _{1}$ takes the
form
\begin{subequations}\label{eq:45}
\begin{eqnarray}\label{eq:45a}
R_{0,1}=R_{0,2}=R_{0,3}=0,\,R_{0,4}=\frac{8}{3}\,\sqrt{\frac{2}{3}}\cos
^{2}\omega _{1}t\,\text{cn}u\,\text{sn}u\,\sin
^{2}\frac{3Jt}{2},\,R_{0,5}=-
\frac{4}{3}\sqrt{\frac{2}{3}}\,\text{cn}u\,\sin
^{2}\frac{3Jt}{2}\sin
2\omega _{1}t, \nonumber\\
R_{0,6}=\frac{2}{9}\sqrt{2}\left( 3\cos 2\omega _{1}t-1\right)
\sin
^{2}\frac{3Jt}{2},\,R_{0,7}=\frac{4}{3}\sqrt{\frac{2}{3}}\,\text{sn}u\,\sin
2\omega _{1}t\sin ^{2}\frac{3Jt}{2}, \nonumber \\
R_{0,8}=\frac{4}{3}\sqrt{\frac{2}{3}}\,\cos ^{2}\omega _{1}t\left(
1-2\text{sn}^{2}u\right) \sin ^{2}\frac{3Jt}{2},
\end{eqnarray}
\begin{eqnarray}\label{eq:45b}
R_{1,1} =\frac{1}{36}\left( 16+12(\cos 3Jt+2)\left(
\text{cn}^{2}u-\text{sn }^{2}u\right) \cos ^{2}\omega _{1}t+2\cos
3Jt-3\cos \left( 3J-2\omega
_{1}\right) t\right. \nonumber\\
\left. -12\cos 2\omega _{1}t-3\cos (3J+2\omega _{1})t\right)
,\,R_{1,2} = \frac{2}{3}(\cos 3Jt+2)\text{cn}u\,\text{sn}u\,\cos
^{2}\omega _{1}t,\,\nonumber\\
\,R_{1,3} =\frac{1}{3}(\cos 3Jt+2)\text{sn}u\,\sin 2\omega
_{1}t,\,R_{1,4}=
\frac{1}{3}\text{sn}u\,\sin 3Jt\sin 2\omega _{1}t,\nonumber\\
\,R_{1,5} =\frac{1}{6}\left( 2\left(
\text{cn}^{2}u-\text{sn}^{2}u\right) \cos ^{2}\ \omega _{1}t+3\cos
2\omega _{1}t-1\right) \sin 3Jt,\,R_{1,6}=
\frac{1}{\sqrt{3}}\,\text{cn}u\,\sin 3Jt\sin 2\omega _{1}t,\nonumber\\
R_{1,7} =-\frac{2}{3}\,\cos ^{2}\omega
_{1}t\,\text{cn}u\,\text{sn}u\,\sin
3Jt,\,R_{1,8}=\frac{1}{\sqrt{3}}R_{1,6},
\end{eqnarray}
\begin{eqnarray} \label{eq:45c}
R_{22} &=&\frac{1}{18}\left( 6(\cos 3Jt+2)\left(
\text{sn}^{2}u-\text{cn} ^{2}u\right) \cos ^{2}\ \omega _{1}t+\cos
3Jt-3(\cos 3Jt+2)\cos 2\omega
_{1}t+8\right) ,\nonumber\\
R_{23} &=&-\frac{1}{3}(\cos 3Jt+2)\text{cn}u\,\sin 2\omega
_{1}t,\,R_{24}=
\frac{1}{\sqrt{3}}R_{16},\,R_{25}=-R_{17},R_{26}=\sqrt{3}R_{14},\nonumber\\
R_{27} &=&\frac{1}{6}\left( 2\left(
\text{cn}^{2}u-\text{sn}^{2}u\right) \cos ^{2}\omega _{1}t-3\cos
2\omega _{1}t+1\right) \sin 3Jt,\,R_{28}=-\frac{1
}{\sqrt{3}}R_{26},
\end{eqnarray}
\begin{eqnarray}\label{eq:45d}
R_{33} &=&\frac{1}{18}\left( -2\cos 3Jt+3\cos \left( 3J-2\omega
_{1}\right) t+12\cos 2\omega _{1}t+3\cos \left( 3J+2\omega
_{1}\right) t+2\right) ,
 \nonumber \\
R_{34} &=&-\frac{2}{3}\,\cos ^{2}\omega _{1}t\left(
\text{cn}^{2}u-\text{sn}
^{2}u\right) \sin 3Jt,\,R_{35}=-R_{14},\,R_{36}=0,\,\nonumber\\
R_{37} &=&-\frac{1}{3}\,\text{cn}u\sin 3Jt\sin 2\omega
_{1}t,\,R_{38}=\frac{4 }{3}\cos ^{2}\omega
_{1}t\,\text{cn}u\,\text{sn}u\sin 3Jt,
\end{eqnarray}
\begin{eqnarray}\label{eq:45e}
R_{44} &=&\frac{1}{72}\left( -72\left( 1-8\text{cn}^{2}u\text{sn}
^{2}u\right) \cos ^{4}\omega _{1}t+8\cos 3Jt-12(2\cos 3Jt+1)\cos
2\omega
_{1}t+9\cos 4\omega _{1}t+19\right) , \nonumber\\
R_{45} &=&\frac{1}{12}\,\text{sn}u\left( 24\left(
\text{sn}^{2}u-3\text{cn} ^{2}u\right) \sin \omega _{1}t\cos
^{3}\omega _{1}t+2(2\cos 3Jt+1)\sin
2\omega _{1}t-3\sin 4\omega _{1}t\right) , \nonumber \\
R_{46} &=&-\frac{2}{3\sqrt{3}}\,\cos ^{2}\omega _{1}t\left( 2\cos
3Jt-9\cos
2\omega _{1}\ t+7\right) \text{cn}u\,\text{sn}u,\nonumber \\
R_{47} &=&\frac{1}{12}\,\text{cn}u\left( -24\left( \text{cn}^{2}u-3\text{sn}%
^{2}u\right) \sin \omega _{1}t\cos ^{3}\omega _{1}t-2(2\cos 3Jt+1)\sin
2\omega _{1}t+3\sin 4\omega _{1}t\right) ,\nonumber \\
R_{48} &=&4\cos ^{4}\omega _{1}t\,\text{cn}u\,\text{sn}u\left(
\text{cn} ^{2}u-\text{sn}^{2}u\right) ,
\end{eqnarray}
\begin{eqnarray}\label{eq:45f}
R_{55} =\frac{1}{18}\left( 6\left( \cos 3Jt+6\cos 2\omega
_{1}t-4\right) \left( \text{sn}^{2}u-\text{cn}^{2}u\right) \cos
^{2}\omega _{1}t-\cos
3Jt+3(\cos 3Jt+2)\cos 2\omega _{1}t\right.\nonumber \\
\quad \left. -9\cos 4\omega _{1}t+1\right) ,\;R_{56}
=\frac{1}{6\sqrt{3}}\, \text{cn}u\left( 4\sin
^{2}\frac{3Jt}{2}\sin 2\omega _{1}t-9\sin 4\omega
_{1}t\right) ,\nonumber\\
R_{57} =\frac{1}{6}\left( 2\cos 3Jt+\cos \left( 3J-2\omega
_{1}\right) t+4\cos 2\omega _{1}t+6\cos 4\omega _{1}t+\cos \left(
3J+2\omega _{1}\right)
t-2\right) \text{cn}u\,\text{sn}u,\nonumber \\
R_{58} =\frac{1}{12}\,\text{cn}u\left( -24\left(
\text{cn}^{2}u-3\text{sn} ^{2}u\right) \sin \omega _{1}t\cos
^{3}\omega _{1}t+2(2\cos 3Jt+1)\sin 2\omega _{1}t-3\sin 4\omega
_{1}t\right) ,
\end{eqnarray}
\begin{eqnarray}\label{eq:45g}
R_{66} &=&\frac{1}{36}\left( -4\cos 3Jt+6\cos \left( 3J-2\omega
_{1}\right) t-12\cos 2\omega _{1}t+27\cos 4\omega _{1}t+6\cos
\left( 3J+2\omega
_{1}\right) t+13\right) ,\nonumber \\
R_{67} &=&\frac{1}{6\sqrt{3}}\,\text{sn}u\left( 2(\cos 3Jt-1)\sin
2\omega
_{1}t+9\sin 4\omega _{1}t\right) ,\,\nonumber\\
R_{68} &=&\frac{1}{3\sqrt{3}}\,\cos ^{2}\omega _{1}t\left( -2\cos
3Jt+9\cos 2\omega _{1}\ t-7\right) \left(
\text{cn}^{2}u-\text{sn}^{2}u\right) ,
\end{eqnarray}
\begin{eqnarray}\label{eq:45h}
R_{77} &=&\frac{1}{18}\left( (\cos 3Jt-1)\left(
3\text{cn}^{2}u-3\text{sn} ^{2}u-1\right) +9\cos 4\omega
_{1}t\left( \text{cn}^{2}u-\text{sn}
^{2}u-1\right) \right.\nonumber\\
&&\quad \left. +3(\cos 3Jt+2)\cos 2\omega _{1}t\left( \text{cn}^{2}u-\text{sn%
}^{2}u+1\right) \right) ,\nonumber\\
R_{78} &=&\frac{1}{6}\,\text{sn}u\left( 6\left(
3\text{cn}^{2}u-\text{sn} ^{2}u\right) \cos ^{2}\omega _{1}t+2\cos
3Jt-3\cos 2\omega _{1}t+1\right) \sin 2\omega _{1}t,
\end{eqnarray}
\begin{equation}\label{eq:45i}
R_{88}=\frac{1}{72}\left( 72\left( 1-8\,\text{cn}^{2}u\,\text{sn}
^{2}u\right) \cos ^{4}\omega _{1}t+8\cos 3Jt-12(2\cos 3Jt+1)\cos
2\omega _{1}t+9\cos 4\omega _{1}t+19\right) ,
\end{equation}
where $u=(ht|k).$ It is straightforward to find the analytic solution for
larger number of qutrits at the same conditions.
\bibliographystyle{unsrt}
\bibliography{Dreferences}
\end{subequations}
\end{document}